\documentstyle[psfig,aps,multicol]{revtex}

\begin{document}
\draft
\title{Dissipative properties of vibrated granular materials}
\author{Clara Salue\~na$^1$, Thorsten P\"oschel$^2$, and Sergei E. Esipov$^3$}
\address{$^1$Departament de F\'{\i}sica Fonamental,
Divisi\'o de Ci\`encies Experimentals i Matem\`atiques,\\
Universitat de Barcelona, Spain \\
$^2$Humboldt-Universit\"at zu Berlin, 
Institut f\"ur Physik, \\ Invalidenstra\ss e 110, D-10115 Berlin, Germany,\\
$^3$30 Newport Pkwy. 1705, Jersey City NJ 07310, U.S.}
\date{February 14, 1998}
\maketitle
\begin{abstract}
  We investigate collective dissipative properties of vibrated
  granular materials by means of molecular dynamics simulations.
  Rates of energy losses indicate three different regimes or ``phases'' in the
  amplitude-frequency plane of the external forcing, namely, solid,
  convective, and gas-like regimes. The behavior of effective
  damping decrement in the solid regime is glassy. 
  Practical applications are dicussed.
\end{abstract}

\begin{multicols}{2}
The dominating approach in the world of vibration control and
suppression by granular systems has been mainly practical \cite{Grubin56}. 
Granular motion relaxes
rapidly once the energy supply is switched off, and
dampers can efficiently absorb energy
released by shocks of external forcing.
Engineers classify granular dampers as {\em passive} ones.

For the case of ``granular gases'', i.e. particulate systems in a
state where the mean free path is large as compared with particle
sizes the cooling rate, the dissipation rate of the system has
been investigated~\cite{Goldhirsch}, and applications
of this work require an analysis of granular gas 
(hydro)dynamics in a given experimental setup. 
Damping in dense granular arrangements is a much more difficult problem
which is mostly studied experimentally.
 
\begin{figure}[htbp]
\begin{minipage}{8cm}
  \centerline{\psfig{file=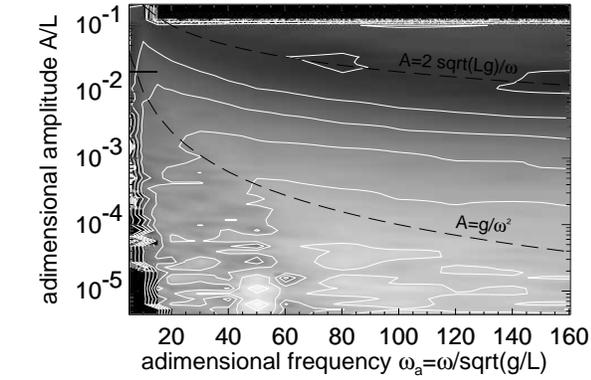,width=8cm,angle=0}}\vspace{1cm}
  \centerline{\psfig{file=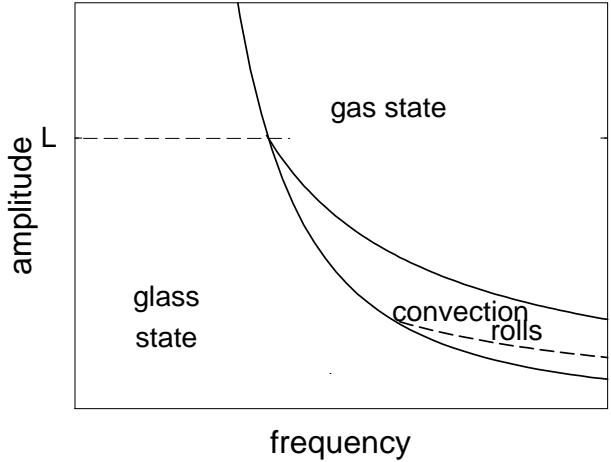,width=8cm,angle=0}}
  \vspace{1cm}
  \caption{Phase diagram of our system,
    and the damping regimes. (a): explicitly analyzed regions,
    the value of the damping parameter $b$ shown by grey scale 
    vs the adimensional frequency and amplitude of shaking
    (white - highest damping). ``Mountains'' at low amplitudes are
    caused by fluctuations in the glassy state. The black stripe at the very
    top of the picture represents empty data and should not be taken into 
    account. The diagram (b) is a summary of regimes based
    on the entire analysis. The narrow region between the glass
    state and dashed line is found to be a non-glassy ``solid''.}
  \label{fig:Phases}
\end{minipage}
\end{figure}

In this work, by using molecular dynamics simulations, we show that
granular systems reveal different damping regimes indicating 
{\em collective} dissipation modes. Our study of these regimes leads 
to a ``phase diagram'' of horizontally vibrated granular 
systems (see Fig.~\ref{fig:Phases}). By using this diagram along with 
the presented
estimates for damping decrements, practitioners may accelerate the
design and testing procedures.

In simulations we focus on two dimensional containers which are partially 
filled with granular material and shaken horizontally. 
The motion of the container is sinusoidal, 
$x(t) = A \sin (\omega t)$; it mimics practical situations where
dampers are tested in the vicinity of the eigenmodes of the vibrating 
mechanism. We
study the reaction of the system to the choice of parameters of
shaking $A$ and $\omega$, keeping all other parameters 
(size, roughness and hardness of particles, filling factor,
size and shape of the apparatus) fixed~\cite{parameters}. 

Our primary objective is the rate of energy dissipation, 
computed by cycle averaging under
steady conditions of oscillatory motion. 
Dissipation is obtained by using two
different ways to ensure consistency of the data: 
(i) from the total power transmitted to the container walls, and (ii) 
from the dissipative work in inter-particle collisions. 
Numerically, both results coincide within a few percent. 

For the molecular dynamics simulations, we use a modified soft-particle
model by Cundall and Strack~\cite{CundallStrack:1979}: Two particles
$i$ and $j$, with radii $R_i$ and $R_j$ and position vectors $\vec{r}_i$
and $\vec{r}_j$, interact if their compression $\xi_{ij}=
R_i+R_j-\left|\vec{r}_i -\vec{r}_j\right|$ is positive. In this case
the colliding spheres feel forces $F_{ij}^{N}$~\cite{KuwabaraKono} and 
$F_{ij}^{S}$~\cite{HaffWerner}, in normal
and shear directions denoted by unit vectors $\vec{n}^N$ and $\vec{n}^S$,
respectively,
\begin{eqnarray}
 F_{ij}^N = \left(Y\sqrt{R^{\,\mbox{\it\footnotesize\it eff}}_{ij}}\right)/
\left( 1-\nu ^2\right) ~\left(\frac{2}{3}\xi^{3/2} + a \sqrt{\xi}\, 
\dot {\xi} \right) \label{normal}\\
F_{ij}^S = \mbox{sign}\left({v}_{ij}^{\,\mbox{\it\footnotesize\it rel}}\right) 
\min \left\{\gamma_S m_{ij}^{\,\mbox{\it\footnotesize\it eff}} 
\left|{v}_{ij}^{\,\mbox{\it\footnotesize\it rel}}\right|~,~\mu 
\left|F_{ij}^N\right| \right\} 
\label{shear}      
\end{eqnarray}
and the resulting momenta acting upon the particles are
$M_i = F_{ij}^S R_i$, $M_j= -F_{ij}^S R_j$. The constant $a$
is the characteristic dissipation rate of the material,
$Y$ is the Young modulus and $\nu$ the Poisson ratio.
The normal and shear friction coefficients, $\gamma _{n} \equiv aY/(1-\nu ^2)$
and $\gamma _{S}$, model dissipation during particle contact. 
Eq.~(\ref{shear}) takes into account that the particles slide upon
each other for the case that the Coulomb condition $\mu \!
\left| F_{ij}^N \right|<\left| F_{ij}^S \right|$ holds, otherwise they feel
some viscous friction. $R^{\,\mbox{\it\footnotesize\it eff}}_{ij} = 
R_iR_j/\left(R_i + R_j\right)$ is the effective radius, and
the effective mass $m_{ij}^{\,\mbox{\it\footnotesize\it eff}}$
is defined analogously. The relative velocity at the point of contact is
\begin{equation}
{v}_{ij}^{\,\mbox{\it\footnotesize\it rel}} = (\dot{\vec{r}}_i - 
\dot{\vec{r}}_j)\cdot \vec{n}^S + R_i {\Omega}_i + R_j {\Omega}_j \; ,
\label{eq:relvel}
\end{equation}
with $\Omega_i$ and $\Omega_j$ being the angular velocities of the particles.

The values of the coefficients used in simulations are $Y/(1-\nu
^2)=7.5\times 10^{7}$, $\gamma _{n}=7\times 10^{2}$, $\gamma _{S}=30$,
$ \mu =0.5$. Cgs units are implied throughout the paper.

\begin{figure}[htbp]
\begin{minipage}{16cm}
  \centerline{\psfig{file=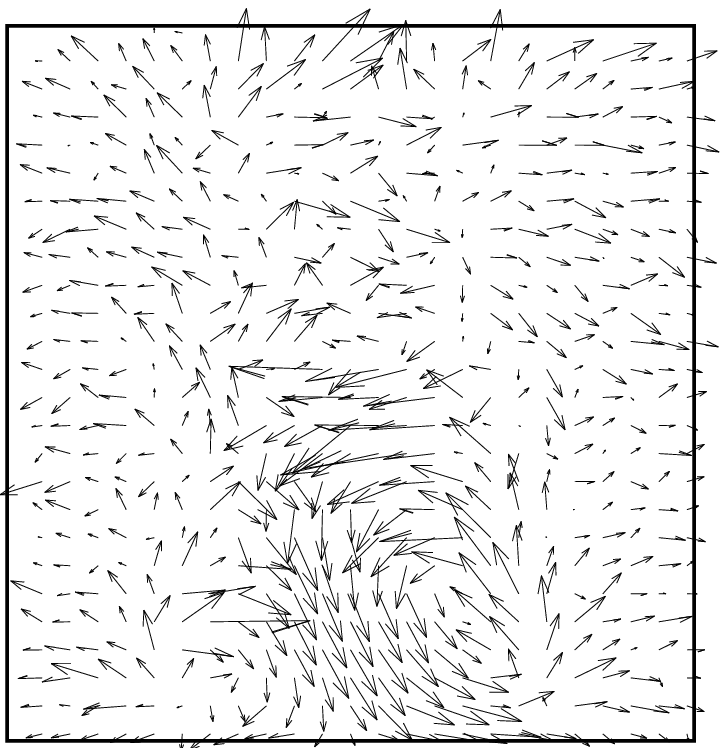,width=5.4cm,angle=0,clip=}\hspace{0.1cm}\psfig{file=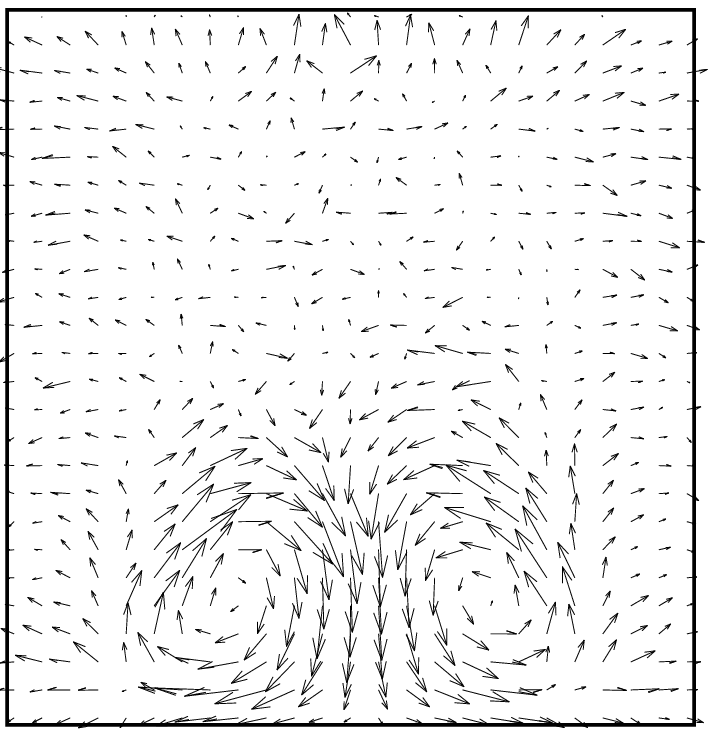,width=5.4cm,angle=0,clip=}\hspace{0.1cm}\psfig{file=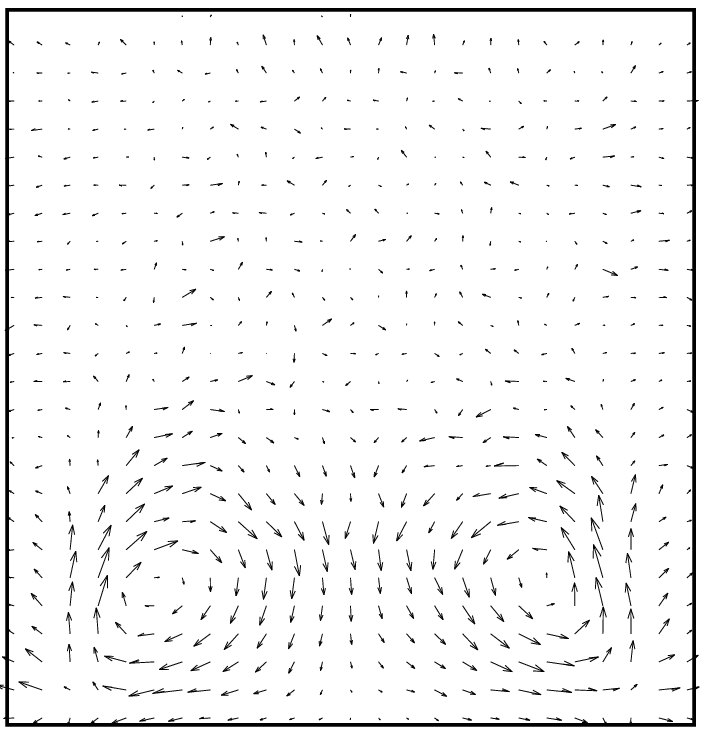,width=5.4cm,angle=0,clip=}}
\vspace{0.1cm}
  \centerline{\psfig{file=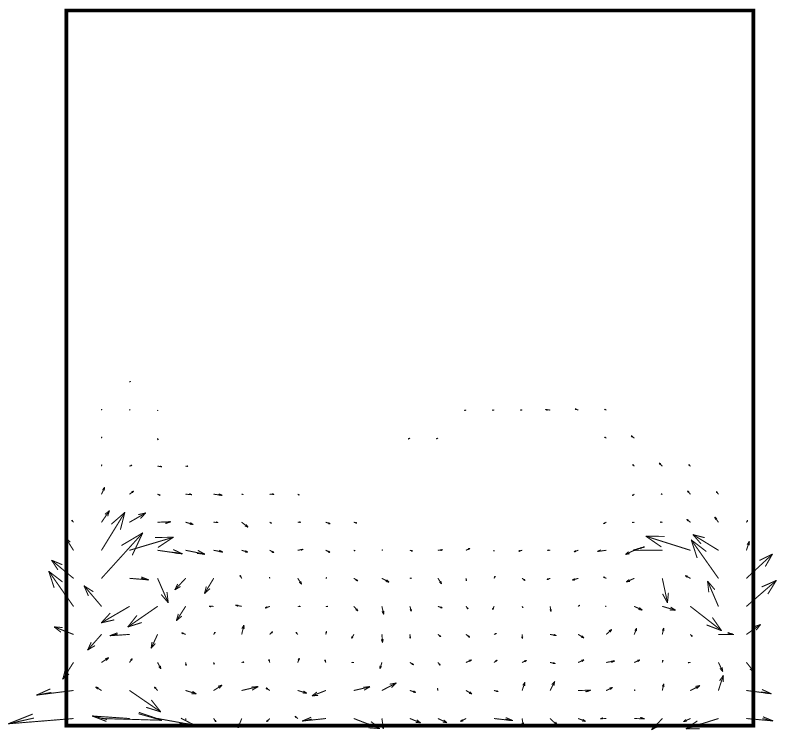,width=5.4cm,angle=0,clip=}\hspace{0.1cm}\psfig{file=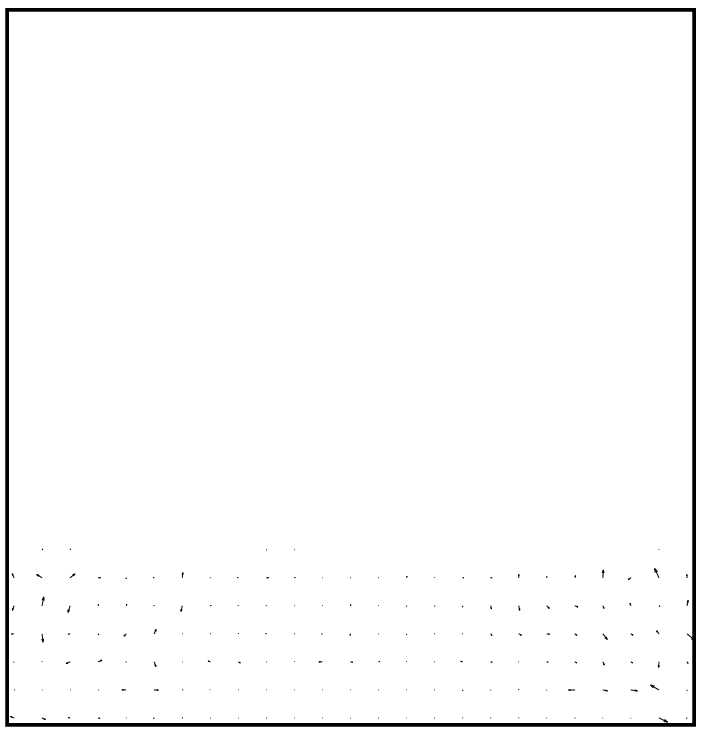,width=5.4cm,angle=0,clip=}}
\vspace{1cm}
  \caption{Images of the cycle-averaged motion of the system vibrated 
    at $f= 20$ for different velocities of forcing, $A\omega$: a)
    1900, b) 1250, c) 600, d) 100, e) 15.}
  \label{fig:cycle}
\end{minipage}
\end{figure}

With the system parameters specified above, depending on forcing one
may find intensive convection (Fig.~\ref{fig:cycle}). Convection 
patterns in horizontally vibrated granular material have
been recently reported~\cite{Metcalfe97}. In our case only two rolls
could be observed. Different aspect ratios or material parameters
give different convection patterns, with 2, 4\cite{Metcalfe97} or more
convection rolls (not shown here)\cite{sandwebpage}.
\end{multicols}
\begin{multicols}{2}

\begin{figure}[htbp]
\begin{minipage}{8cm}
    \centerline{\psfig{file=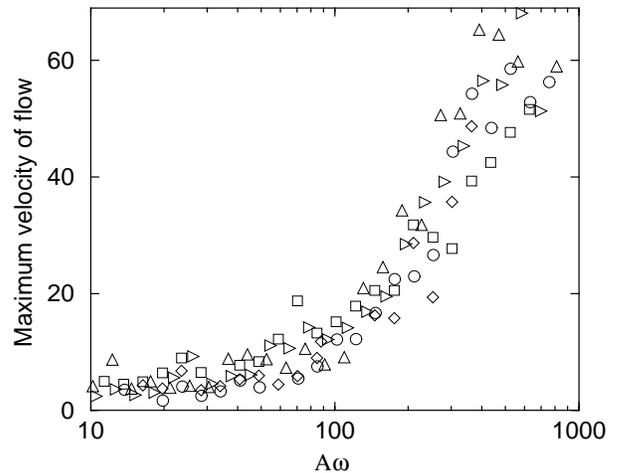,width=8cm}}
\vspace{1cm}
  \caption{Maximum velocity of collective flow $V$ over amplitude 
    of shaking velocity $A\omega$, for the different frequencies
    analyzed: $\diamond: 5$; $\Box: 10$; $\circ: 20 $;
    $\triangleright: 40 $; $\triangle: 80$ (Hz).}
  \label{fig:prova}
\end{minipage}
\end{figure}

The velocity profiles in Fig.~\ref{fig:cycle} have been produced by
averaging particle velocities in the regime of steady oscillations. We do not intend
to discuss the effect of convection in horizontally shaken material in
detail, although we note that the onset of convection in the
system is due to a critical amplitude of driving velocity
$\left(A\omega\right)_{cr}$.  Fig.~\ref{fig:prova} shows the maximum
absolute value of convective motion in the system, i.e. the length of
the longest arrow in Fig.~\ref{fig:cycle}, over the velocity amplitude
$A\omega$ for different frequencies $f$ (in Hz). Below
$\left(A\omega\right)_{cr}\approx 60$ there is slow collective
motion in the system, but close to this point there is a transition
into a rapid convective regime (see also Fig.~\ref{fig:cycle}d).

To characterize the dissipation in the system we introduce an
effective damping parameter $b$ which is proportional to the ratio
between the averaged dissipated power per cycle $T$ and the mean
translational kinetic energy of the granular system~\cite{rot},
\begin{equation}
b=\frac{\frac{1}{T}\int_{T}W_{\mbox{\footnotesize\em diss}}(t)dt}
{2\sum_n m_n \int_T v_{n}^{2}(t) dt}\,.
\label{b}
\end{equation}
Here summation is performed over all granular particles in the container. 
In the reference case of linear oscillator 
the damping decrement $b$ equals the inverse of the amplitude
relaxation time.

\begin{figure}[htbp]
\begin{minipage}{8cm}
  \centerline{\psfig{file=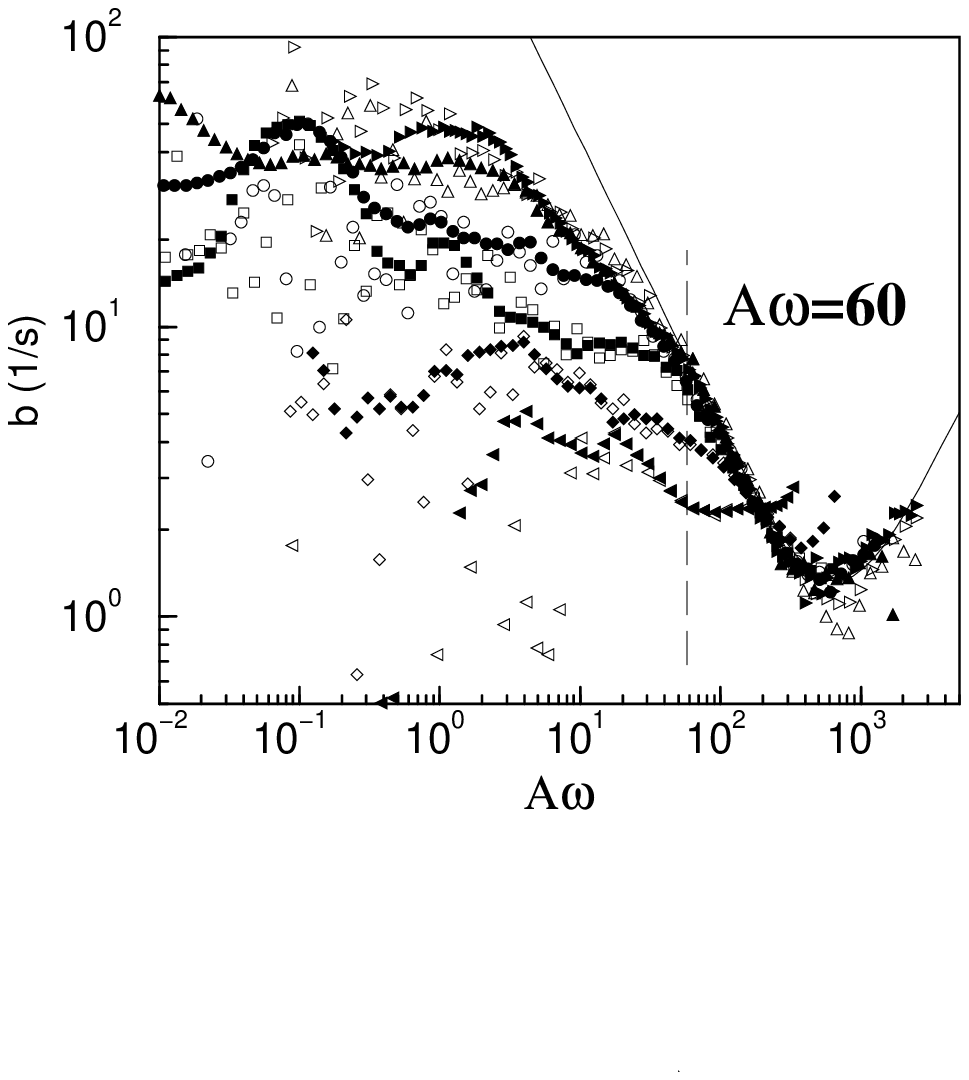,width=8cm,bbllx=27pt,bblly=175pt,bburx=310pt,bbury=420pt,angle=0}}\vspace{1cm}
%  \vspace{1cm}
  \centerline{\psfig{file=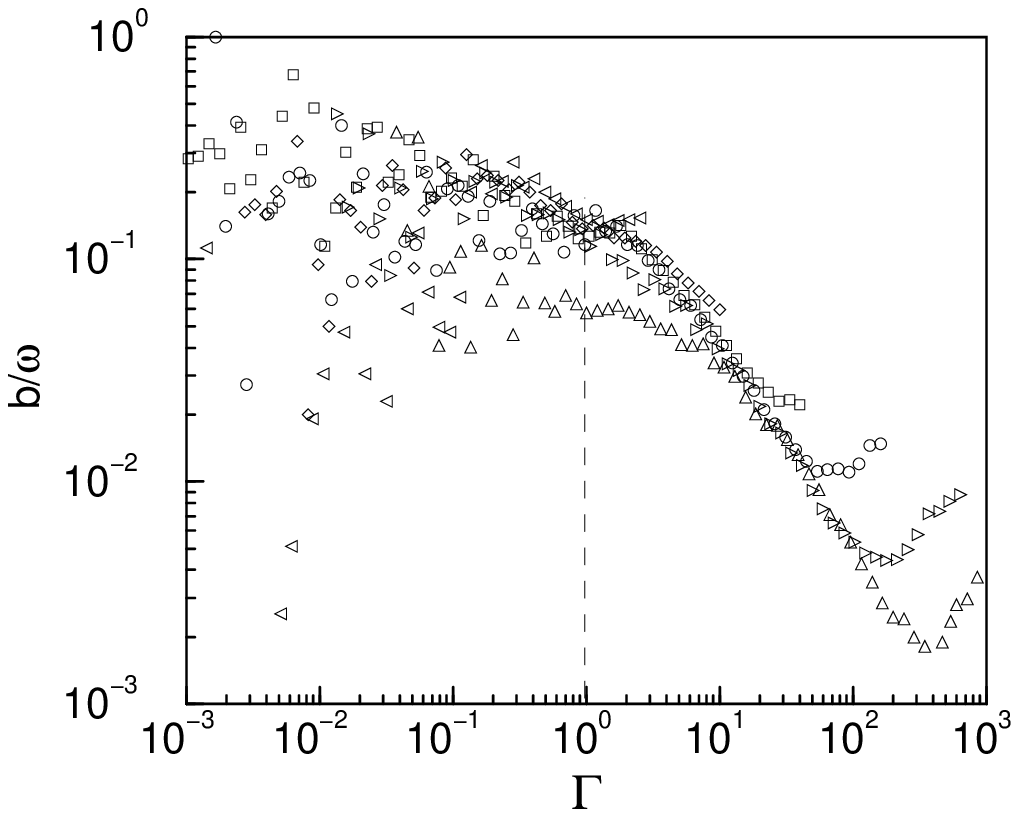,width=8cm,bbllx=27pt,bblly=175pt,bburx=310pt,bbury=420pt,angle=0}}
%  \vspace{2cm}
  \caption{Effective damping parameter $b$ vs the amplitude of 
    velocity of vibration, $A\omega$ (a); vs the dimensionless forcing acceleration
    parameter, $\Gamma$ (b). Different symbols refer to different
    frequencies; $\triangleleft: 2.5$; $\diamond: 5$; $\Box: 10$;
    $\circ: 20 $; $\triangleright: 40 $; $\triangle: 80$ (Hz). Open symbols
    refer to fast cooling schedule, filled symbols refer to slow cooling
    (see text for details). The solid line in (a) shows a fit using
    (\ref{bgas}) and (\ref{bliq}). Observe the fit lines with 
    slopes -1 and +1 for $b \propto (A\omega)^{-1}$ and 
    $b \propto (A\omega)^{+1} $in the ``liquid'' and ``gas'' regimes, resp.}
  \label{fig:effDamping}
\end{minipage}
\end{figure}

Fig.~\ref{fig:effDamping} shows the damping $b$ as a function of the
effective acceleration $\Gamma=A\omega^{2}/g$ and the amplitude of the
velocity of vibration, $A\omega$. Different symbols (filled or open)
display different frequencies. Except for the very low frequency
range, $b$ scales with the amplitude of the velocity of vibration,
$A\omega$ (Fig.~\ref{fig:effDamping}a). A transition from the
$A\omega$-scaling into another regime takes place at $\Gamma\approx 1$
(Fig.~\ref{fig:effDamping}b). Below this point the effective
damping parameter becomes less sensitive to $\Gamma$ and fluctuates
very strongly.

Fig.~\ref{fig:effDamping} reveals three regions, indicative of
different dynamic regimes. The local minimum 
in Fig.~\ref{fig:effDamping}a is the region of spatially
organized behavior with well developed convection rolls, where the
entire granular mass participates in collective motion
(Fig.~\ref{fig:cycle}c). For higher velocity amplitudes $A\omega$, 
the dissipation rate increases, as
particles begin to fly across the box and the ordered structure of the
rolls starts to disappear (cf. Fig.~\ref{fig:cycle}a-b). The minimum
corresponds to a characteristic velocity at which the system begins to 
display the ``gas'' state. This characteristic velocity 
can be estimated by equating the characteristic time of motion in horizontal and
vertical directions for a particle with velocity of the order of
$A\omega$,
\begin{equation}
L/A\omega = A\omega/g\,.
\label{cond1}  
\end{equation}
For $L = 100$, this gives $A\omega = 313$, which gives an 
order-of-magnitude estimate of 
the minimum of Fig.~\ref{fig:effDamping}a. Additional analysis
reveals that the correct prefactor is close to 2 (see 
Fig.~\ref{fig:Phases}a and discussion below). At higher velocities
particles continue to stay airborne, and gravity becomes unimportant.
Thus, the boundary $A\omega \sim \sqrt{Lg}$ separates the ``liquid'' and
gas-like regimes~\cite{Lreskalierung}.

To the left from the minimum, with amplitude decrease at any
frequency, the depth of the rolls diminishes progressively until a
critical value $(A\omega)_{cr}\approx 60$ is reached. At
this velocity the rolls vanish completely and we do not find organized
motion in the system anymore (cf. Figs.~\ref{fig:prova} and
\ref{fig:cycle}d). The system seems to be in a ``solid'' state. 
The critical value of velocity is manifested in
Fig.~\ref{fig:effDamping}a by a change in the damping slope. Given that the
acceleration of shaking here exceeds $g$, critical velocity in this 
range points to $A\omega \sim \sqrt{Rg}$, where $R$ is the particle
radius. This is consistent with our particle sizes~\cite{parameters}. 

In the solid regime one has to switch from velocities to accelerations.
Following Fig.~\ref{fig:effDamping}b to small
accelerations, one finds a change in the behavior at
$\Gamma\approx 1$, i.e. when the maximal acceleration of shaking
becomes comparable with gravity $g$. For $\Gamma\gtrsim 1$ the curves
are smooth, but for $\Gamma\lesssim 1$ suddenly the data become very
noisy. At this point the grains form a glassy ``solid'' phase. We note that
our understanding of the transition at $\Gamma\lesssim 1$ is different from the 
solid-fluid transition reported recently~\cite{Ristow&al97} as long as 
$A\omega \lesssim \sqrt{Rg}$. At higher velocities our results lead to the
same conclusions as the results reported in Ref.~\cite{Ristow&al97}.

To reiterate, the transition point found in
Ref.~\cite{Ristow&al97} is given by $A\omega^2 = g$ in our notation. 
It is unclear whether the non-glassy solid
phase at $\Gamma\gtrsim 1$ can be identified by the method of granular temperature used in
Ref.~\cite{Ristow&al97}. If our understanding is correct, the non-glassy 
regime is not necessarily fluidized, and the region of increasing granular temperature
may overlap the region of non-vanishing configurational order 
associated with a ``solid'' state. This is an intriguing issue, and
it is worth a separate study. 

At the glass transition point, $\Gamma\approx 1$, particle {\it ensembles} get
trapped in their momentary position. In the glassy regime the rate of
dissipated energy is strongly sensitive to the configurational arrangement
of the particles i.e. to the history of the system. In other words, depending on initial 
conditions, for the same parameters of oscillation $A$ and $\omega$, 
the system can settle in different configurational arrangements,
and exhibit different dissipation rates. Transitions between different
condensed phases based on multi-particle condensation have been suggested
earlier by two of the authors~\cite{Goldhirsch}.

To demonstrate this fact, in
Fig.~\ref{fig:effDamping}a two different sets of data points
are shown by filled
and open symbols; they refer to ``fast'' and ``slow'' cooling. 
By fast
cooling we mean an instantaneous transition from any fluidized initial
state to the specified amplitude and frequency of vibration. The
initial conditions, namely, random positions and velocities of
particles, are kept the same for each data point. Thus, there is no
memory of the preceding evolution. Having the first 30 cycles of the
driving oscillation discarded, the mean dissipation is obtained by
averaging over 30 cycles of shaking. The
large fluctuation of the dissipation rate $b$ is not due to
insufficient data averaging: averaging over 100 cycles, instead of 30,
leads to the same results. In the case of slow cooling, the system is
initialized only once for each frequency, and the final state for a
certain amplitude $A$ serves as the initial one for the next amplitude
to be investigated. The amplitude of vibration in each simulation is
diminished by a factor of 1.2. The behavior in this case can be sensitive
to the entire history of the experiment. According to the mechanism
discussed in Fig.~\ref{fig:effDamping}a, in the region
$\Gamma\lesssim 1$ the fast cooling data points begin to fluctuate,
whereas the slow cooling data are relatively smooth. Thus, for
accelerations $\Gamma\lesssim 1$ one identifies a glass regime.

The different damping regimes can be clearly appreciated in 
Fig.~\ref{fig:Phases}a, where a shaded contour plot of the 
effective damping parameter is shown as a function of the
adimensional amplitude and frequency of shaking~\cite{Lreskalierung}.
These results suggest that,
on the $(A,\omega)$ plane, the regime boundaries form a diagram similar
to the Van der Waals system, with the ``triple point'' roughly
located at $(L,\sqrt{g/L})$.
Above this point, direct ``sublimation'' is achieved at $A\omega^2
\approx g$. Convection rolls can develop at $A < L$ above a critical
velocity, as has been discussed. This diagram is shown schematically in
Fig.~\ref{fig:Phases}b, where for the seek of clarity the
boundaries between regimes are sharpened.
The damping decrement along with convection dynamics 
provide indicative suggestions about ``phases'' which we discuss here. 
Any sound indentification of transitions and phases 
can only be based on studies of appropriate order parameters.
In principle, it is conceivable that some ``phases'' may 
represent only preferred dynamic states. 

The fluidized regime is amendable to hydrodynamical arguments.
Let us consider the ``gas'' state, where the dissipation is dominated by
collisions with the walls of the container. The pressure transmitted
to the vertical wall by the granular cloud of density $\rho$ which
moves with velocity $v=A\omega$ relative to the wall is $\sim\rho
A^2\omega^2 $. Neglecting other contributions, 
the dissipated power during the collision is then roughly a
fraction of the quantity $\rho v^3 L$. Using Eq.~(\ref{b}) one
finds
\begin{equation}
b=C_g\frac{A\omega}{L}\,.
\label{bgas}
\end{equation}
Formula (\ref{bgas}) is applicable at $A\omega \gtrsim \sqrt{Lg}$.
$C_g$ is some unknown numerical prefactor. This expression
is simply proportional to the collision frequency of independent 
gas particles moving at velocity $A\omega$ and having a mean free 
path of the order of $L$. In other words, collective modes are
irrelevant in the ``gas'' phase as it should be.

For velocities $A\omega < \sqrt{Lg}$, the system is only partially
fluidized. This means that the pressure transmitted by the container
walls does not exceed the static stress that the condensed particles can
sustain. The latter is roughly of the order of $ p =\rho g L$ in horizontal direction. 
As long as there is any fluidized material in the system, these stresses
cannot be lower. Therefore, the dissipated power is $pHv$, where $H$
is the height of the system. Dividing it by the total kinetic energy 
$\frac{1}{2} \rho H L v^2$ one gets
\begin{equation}
b=C_l\frac{g}{A\omega}\,.
\label{bliq}
\end{equation}
Formula (\ref{bliq}) is applicable at $A\omega^2 \gtrsim g $, $A\omega \gtrsim \sqrt{Rg} $, 
$A\omega \lesssim \sqrt{Lg}$, and $C_l$ is the unknown numerical prefactor. 
Very roughly, in this regime some condensed particles transmit their 
motion to other particles and the latter can move against gravity. 
$b$ is the inverse time needed to decelerate them. The
best fit of the curves of Fig.~\ref{fig:effDamping} provides the
values $C_g = 0.1$ and $C_l = 0.45$ for our simulated system, giving
a value for the minimum $(A\omega)_{G} = 2.1 \sqrt{Lg}$. This
is obviously consistent with the upper curve shown in Fig.~\ref{fig:Phases}a,
separating the ``fluid'' and ``gas'' regimes.

To summarize, by means of MD simulations we have shown how the analysis of
energy-loss rate displays different damping regimes. In particular,
one finds that fully convective states correspond to minimal damping decrement. 
Performing two types of measurements, referred as
to slow and fast cooling, one identifies a glass regime. This regime,
in which configurational states affect the dynamical properties of the
system, is separated from the fluidized regime by the value of the
forcing parameter, $\Gamma \equiv A\omega^{2}/g \sim 1$. For higher forcing one
may find a non-glassy solid phase as long as the velocity of shaking is 
smaller than $A\omega \lesssim\sqrt{Rg}$. The
$A\omega$-scaling of the damping curves signals the beginning of the
``fluid'' regime. Here convective states can develop in a region of the
plane $(\omega, A)$ above the critical velocity $A\omega \sim\sqrt{Rg}$, 
critical acceleration $A\omega^{2}/g \gtrsim 1$  and below ``evaporation''
threshold $ A = \sqrt{Lg}/\omega$. The damping decrement passes through the 
minimum for velocities $A\omega \sim
\sqrt{Lg}$, and at higher velocities a gas-like 
state can be identified, where the effects of gravity are negligible.

The authors thank V.~Buchholtz for discussions.  Invaluable help and
encouragement by S.~Simonian and multiple discussions of practical issues 
are gratefully acknowledged. The calculations have been done on the parallel
computer {\it KATJA} (http://summa.physik.hu-berlin.de/KATJA/) of the
medical department {\em Charit\'e} at the Humboldt University in Berlin.
The work was supported in part by Deutsche Forschungsgemeinschaft through Po
472/3-2.

\end{multicols}

\begin{references}
\bibitem{rot} The rotational energy has not been taken into account in
  the denominator of (\ref{b}) because the rotational energy
  represents a small fraction of the translational kinetic energy
  (less than 5\%).
\bibitem{Lreskalierung} Note that the existence of the $L$-dependent
  transition point $A\omega = \sqrt{Lg}$ along with system-independent 
  threshold $\Gamma
  = 1$, makes it necessary to perform a proper
  scaling of amplitudes (by $L$), frequencies (by $\sqrt{g/L}$) and velocities
  (by $\sqrt{Lg}$) in order to compare results
  for different container lengths $L$ in the shaking direction.
\end{references}
\end{document}